\documentclass[sigconf]{acmart}

\usepackage{graphicx}
\usepackage{textcomp}
\usepackage{xcolor}
\usepackage{xspace}
\usepackage{tabularx}
\usepackage{hyperref}
\usepackage{balance}
\usepackage{booktabs}
\usepackage[inline]{enumitem}
\usepackage[justification=centering]{caption}
\usepackage{subcaption}

\newcommand{\etal}{et al.\xspace}
\newcommand{\WoCP}{\mbox{WoC-P}\xspace}

\makeatletter
\newcommand{\linebreakand}{%
  \end{@IEEEauthorhalign}
  \hfill\mbox{}\par
  \mbox{}\hfill\begin{@IEEEauthorhalign}
}
\makeatother

\usepackage[]{todonotes}

\copyrightyear{2022} 
\acmYear{2022} 
\setcopyright{acmcopyright}\acmConference[BotSE 2022]{Fourth International Workshop on Bots in Software Engineering}{May 9, 2022}{Pittsburgh, PA, USA}
\acmBooktitle{Fourth International Workshop on Bots in Software Engineering (BotSE 2022), May 9, 2022, Pittsburgh, PA, USA}
\acmPrice{15.00}
\acmDOI{10.1145/3528228.3528403}
\acmISBN{978-1-4503-9333-1/22/05}

\definecolor{myblue}{HTML}{3276AF}
\definecolor{myorange}{HTML}{F38636}

\begin{document}

\title{Leveraging Predictions From Multiple Repositories to Improve Bot Detection}
\author{Natarajan Chidambaram}
\email{natarajan.chidambaram@umons.ac.be}
\affiliation{%
    \institution{Software Engineering Lab -- University of Mons}
    \streetaddress{Avenue Maistriau 15}
    \city{Mons}
    \country{Belgium}
    \postcode{7000}
}

\author{Alexandre Decan}
\email{alexandre.decan@umons.ac.be}
\affiliation{%
    \institution{Software Engineering Lab -- University of Mons}
    \streetaddress{Avenue Maistriau 15}
    \city{Mons}
    \country{Belgium}
    \postcode{7000}
}

\author{Mehdi Golzadeh}
\email{mehdi.golzadeh@umons.ac.be}
\orcid{0000-0003-1041-439X}
\affiliation{%
  \institution{Software Engineering Lab -- University of Mons}
  \streetaddress{Avenue Maistriau 15}
  \city{Mons}
  \country{Belgium}
  \postcode{7000}
}

\begin{abstract}
Contemporary social coding platforms such as GitHub facilitate collaborative distributed software development.
Developers engaged in these platforms often use machine accounts (bots) for automating effort-intensive or repetitive activities. Determining whether a contributor corresponds to a bot or a human account is important in socio-technical studies, for example to assess the positive and negative impact of using bots, analyse the evolution of bots and their usage, identify top human contributors, and so on.
BoDeGHa is one of the bot detection tools that have been proposed in the literature. It relies on comment activity within a single repository to predict whether an account is driven by a bot or by a human.
This paper presents preliminary results on how the effectiveness of BoDeGHa can be improved by combining the predictions obtained from many repositories at once.
We found that doing this not only increases the number of cases for which a prediction can be made, but that many diverging predictions can be fixed this way.
These promising, albeit preliminary, results suggest that the ``wisdom of the crowd'' principle can improve the effectiveness of bot detection tools.
\end{abstract}

\maketitle

\section{Introduction}
\label{sec:intro}

Social coding platforms like GitHub promote collaboration and interaction between contributors \cite{software-social}. Along with this opportunity for engagement, developers also face some workload in performing error-prone and repetitive tasks such as conducting regular dependency checks, deploying release, testing code, reviewing code, merging pull requests and so on \cite{power_of_bots}. While small projects can thrive under the guidance of a lone developer \cite{contributions_count}, the complexity and workload of larger projects makes it difficult to keep up with the pace of maintaining high-quality software releases. To reduce this workload and to automate effort-intensive and repetitive activities, automated tools/bots (machine accounts that works without human intervention) are used~\cite{Storey2016, Lebeuf2017a, Lin2016, Wessel2018, Farooq2016, Urli2018, Beschastnikh2017}.
These bots might also influence the software development process, either positively or negatively, depending on the role that they are assigned and the way that they are being used~\cite{bot_desruptive}.

Identifying the presence of these bots is not only useful for researchers conducting socio-technical studies but also for practitioners and funding organizations to identify contributors and to accredit them.
The research literature already lists a few approaches to identify bots in software repositories, such as BIMAN \cite{BIMAN}, BoDeGHa \cite{Bodegha} or BoDeGic~\cite{Bodegic}.
BIMAN~\cite{BIMAN} combines three different approaches to recognize bots in commits: (i)~the presence of the string ``bot'' at the end of the author name, (ii)~repetitive commit messages, and (iii)~features related to files changed in commits.
BoDeGHa~\cite{Bodegha} analyses comments posted in issues and pull requests to detect bots, based on the assumption that bots tend to frequently use a limited set of comment patterns.
BoDeGic~\cite{Bodegic} transposes this approach to commit messages, assuming that bots tend to have a limited set of commit message patterns.
Golzadeh \etal~\cite{GolzadehBotse2021} proposed a probabilistic model based on NLP techniques to detect bot activity at the level of individual comment in issues and pull requests.
In a recent work~\cite{botse_paper_2}, they also compared 5 different approaches to detect bots in software repositories, including the above ones, and found that none of them is accurate enough to capture all bots.

Our goal is to improve the detection of bots active in issue and pull request comments, that is, to improve BoDeGHa.
BoDeGHa is a tool that, given the name of a GitHub repository, predicts for each contributor with enough activity in the repository whether this contributor corresponds to a \emph{bot} or a \emph{human} contributor. If a contributor has not made enough comments, BoDeGHa classifies it as \emph{unknown}.

Although BoDeGHa has been shown to perform well in detecting bots~\cite{Bodegha}, it may still wrongly classify some contributors.
Because BoDeGHa works at the repository level, this means that a same contributor active in multiple repositories may lead to diverging predictions, this is, it may be classified as \emph{bot} in some repositories and as \emph{human} in some other ones.
For example, while BoDeGHa identifies the well-known \emph{dependabot} bot correctly in many different repositories, it identifies it as a \emph{human} contributor in \emph{artichoke/rand\_mt} because the 24 comments made by \emph{dependabot} in this repository exhibit 10 different comment patterns, corresponding to the behaviour usually observed for human contributors.
At the same time, BoDeGHa classifies the same bot as \emph{unknown} in \emph{cossacklabs/themis} because it only has 9 comments in this repository.
Similarly, a human contributor can be sometimes classified as a bot. For example, in the GitHub repository \emph{rust-lang/libc} we found a human contributor\footnote{Name is hidden to comply with GDPR regulations.} that is detected as \emph{bot} because most of his/her comments follow a single comment pattern of the form ``\emph{bors r+}''. On the other hand, this contributor is correctly classified as \emph{human} in \emph{crossbeam-rs/crossbeam} and \emph{rust-lang/rust} for example.

In this paper, we investigate how frequently such situations occur in GitHub repositories. We quantify how frequently do contributors have diverging predictions (that is, predicted as \emph{bot} and \emph{human} by BoDeGHa), and how frequently they have incomplete predictions (that is, predicted as \emph{unknown} by BoDeGHa). We provide preliminary insights on a novel approach to improve the accuracy of BoDeGHa by leveraging predictions from multiple repositories. We evaluate to which extent diverging and incomplete predictions can be fixed based on the wisdom of the crowd principle.
More specifically, we address the following research questions:

\noindent \emph{RQ0: How frequently are contributors active in multiple repositories?} We observe that one third of the contributors are active in multiple repositories.\\
\emph{RQ1: How frequently do contributors have diverging or incomplete predictions?} More than half of the contributors identified at least once as bots have diverging or incomplete predictions.\\
\emph{RQ2: To which extent can we fix diverging predictions?} We show that an approach based on the wisdom of the crowd principle is effective at fixing diverging predictions.\\
\emph{RQ3: To which extent can we complete predictions?} We show that the same approach is promising to address incomplete predictions.\section{Dataset}
\label{sec:Dataset}

BoDeGHa bot identification tool takes as input a GitHub repository and outputs whether the contributors in this repository correspond to bot or human contributors.
Since our goal is to improve the performance of BoDeGHa by leveraging predictions from multiple repositories, we need a large collection of GitHub repositories having their contributors active in multiple repositories.
Following the advice of Kalliamvakou et al.~\cite{Kalliamvakou2014} we need to avoid repositories that have been created merely for experimental or personal reasons, or that only show sporadic traces of activity. Good candidate datasets are collections of repositories associated to the collaborative development of open source software packages for specific programming languages.

We collected the GitHub repositories associated with the software packages that are distributed through the \textsf{Cargo} package manager, for the Rust programming language.
In October 2021, 68,621 Rust packages were available on \textsf{Cargo} and 38,886 of them (i.e., 56.7\%) have an associated repository on GitHub.
Since we need bots to be active in the repositories to conduct our empirical study, and since bots are more likely to be present in larger and more mature projects, we excluded packages that do not even refer to their homepage or to their documentation.
This left us with 22,156 packages.
Given that BoDeGHa relies on the comments made in issues and pull requests to identify bot contributors, we excluded repositories having less than 100 issues or pull requests.
At the end of the data extraction process, the dataset contains 1,039 GitHub repositories accounting for 147,426 pairs of contributor/repository.\section{Findings}
\label{sec:findings}
\subsection*{RQ0: How frequently are contributors active in multiple repositories?}

Since we aim to improve bot detection by leveraging predictions made on multiple repositories, we need contributors to be active in more than a single repository.
This question aims to quantify how frequently contributors are active in multiple repositories.
The 147,426 pairs of contributor/repository in our dataset correspond to 57,757 distinct GitHub accounts, already indicating that some contributors are active in more than one repository. Only 8,532 contributors out of these 57K (14.8\%) have enough commenting activity in at least one repository for BoDeGHa to be applied.
For each of these 8,532 contributors (i.e., each distinct GitHub account), we counted the number of repositories that each contributor was active in.
Table~\ref{table:repoprop} reports on the number and proportion of contributors in function of the number of repositories they are active in.

\begin{table}[t]
	\centering
	\caption{Number and proportion of contributors in function of the number of repositories they are active in}
	\label{table:repoprop}
	\begin{tabular}{ r | r  r  r  r  r  r }
		 \# repositories $\rightarrow$ & 1 &  2 & 3 & 4 or 5 & 6 - 9 & 10+\\
		\hline
		\# contributors & 5,671 & 1,530 & 496 & 385 & 239 & 211\\
		\% contributors & 66.5\% & 17.9\% & 5.8\% & 4.5\% & 2.8\% & 2.5\%\\
	\end{tabular}
\end{table}

We observe that while most contributors (5,671 out of 8,532, 66.5\%) are active in a single repository only, around one third of the contributors (2,861, i.e., 33.5\%) are active in multiple repositories.
We will focus on those 2,861 contributors since they correspond to those for which BoDeGHa will produce several, potentially diverging (i.e., \emph{bot} and \emph{human}) or incomplete (i.e., \emph{unknown}) predictions.
These 2,861 contributors are active in a total of 1,010 distinct repositories.

\subsection*{RQ1: How frequently do contributors have diverging or incomplete predictions?}

We applied BoDeGHa on each of the 1,010 repositories identified in $RQ0$ in order to get the predictions for each of the 2,861 contributors active in two or more repositories.
Under the hood, BoDeGHa downloads up to 100 pull request or issue comments for each contributor active in the repository. Only the comments made during the last five years (i.e., after December 2016) are considered. BoDeGHa then analyses these comments and predicts whether the contributor corresponds to a \emph{bot} or a \emph{human} contributor based on several features including the repetitiveness of comments and the number of comment patterns. If a contributor has less than 10 comments, BoDeGHa classifies it as \emph{unknown}.
At the end of this process, we have a total of 41,542 predictions of which 1,146 correspond to \emph{bot}, 10,227 to \emph{human} and 30,169 to \emph{unknown}.
The high proportion of \emph{unknown} predictions (73\%) indicates that most contributors have less than 10 comments in the considered repositories.

Since our focus is on improving bot detection, we select contributors that were classified \emph{bot} at least once.
Out of the initial 2,861 distinct contributors active in at least two repositories, 229 (8\%) were classified \emph{bot} at least once. Among them, 106 (46\%) were consistently classified \emph{bot} in all the repositories they were active in. Out of the 123 remaining contributors having been predicted as \emph{bot} at least once, 60 have diverging predictions (i.e., they were also classified as \emph{human}) and 63 have consistent but incomplete predictions (i.e., they were also classified as \emph{unknown}).

To assess to which extent bot detection can be improved by leveraging predictions from multiple repositories, we need to determine the correct type (i.e., bot or human) of each account.
Two authors of this paper manually and independently checked the 3,086 predictions for the 229 contributors that were at least once predicted as bot to determine their actual type, following an inter-rater agreement process.
The first step of this process ended up with an agreement on 95\% of the cases. The remaining ones were discussed together, ending up with an agreement on all of them.
With this process, we found that BoDeGHa incorrectly predicted \emph{bot} in 110 cases and incorrectly predicted \emph{human} in 31 cases.
Table~\ref{table:classsummary} summarizes the number of actual bot and human contributors we found, as well as the number of \emph{bot}, \emph{human} and \emph{unknown} predictions obtained for them.

\begin{table}[!h]
	\centering
	\caption{Number of actual bot and human contributors, and their number of \emph{bot}, \emph{human} and \emph{unknown} predictions}
	\label{table:classsummary}
	\begin{tabular}{ r | r | r  r  r }
			       &  & \multicolumn{3}{c}{predictions}\\
		 & contributors & \# \emph{bot}          & \# \emph{human}   & \# \emph{unknown}\\
		\hline
		actual bot & 142 & 1,110 & 31 & 413\\
		actual human & 87 & 110 & 288 & 1,134\\\hline
		total & 229 & 1,220 & 319 & 1,547\\
	\end{tabular}
\end{table}
\vspace{-0.5cm}
\subsection*{RQ2: To which extent can we fix diverging predictions?}

\begin{figure}[!h]
	\centering
	\includegraphics[width = \columnwidth]{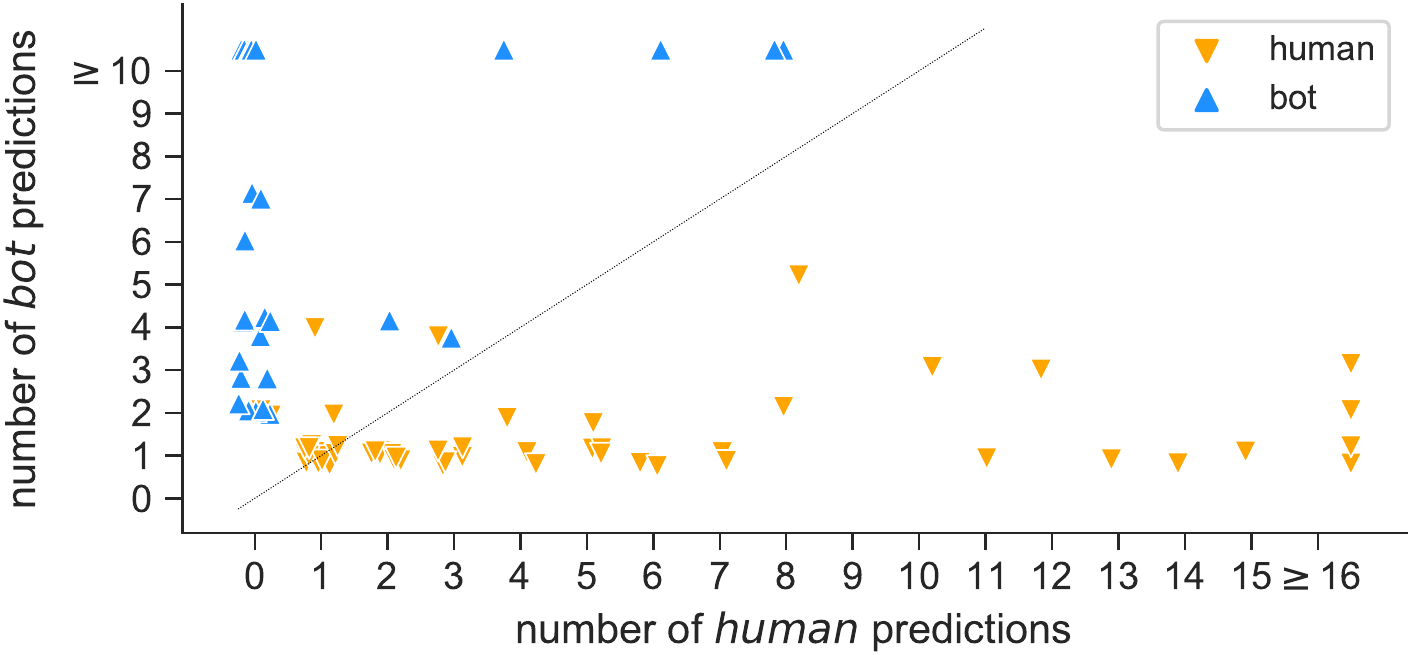}
	\caption{Number of \emph{bot} and \emph{human} predictions, each point is a contributor}
	\label{fig:bothum}
\end{figure}

Previous research question revealed that many contributors have different predictions depending on the repository BoDeGHa is applied on.
In this research question, we propose an approach based on the wisdom of the crowd principle to fix these diverging predictions.
More specifically, if one assumes that BoDeGHa is more often correct than wrong in predictions, then, given a contributor having multiple predictions, we can assume that the most frequent prediction (either \emph{bot} or \emph{human}) is correct, while the less frequent one is not.
Let \WoCP be such bot detection model.
\WoCP stands for \emph{Wisdom of the Crowd principle for Predictions} and works on top of BoDeGHa by automatically replacing the less frequent predictions of a contributor with the most frequent ones. Ties are arbitrarily resolved as \emph{human}.
We applied both BoDeGHa and \WoCP on the 84 contributors that have at least two predictions of which one is \emph{bot}.
Figure~\ref{fig:bothum} shows, for each contributor, the number of \emph{human} predictions, the number of \emph{bot} predictions, and whether it is an actual bot or human. To permit distinguish overlapping points, we added a jitter of 0.25 on both axes. The diagonal line illustrates the \WoCP model: any contributor above the line will be consistently predicted as a bot (i.e., the \emph{human} predictions are replaced by \emph{bot} predictions), while any contributor below will be consistently predicted as a human (i.e., the \emph{bot} predictions are replaced by \emph{human} predictions).

As can be observed from the figure, the approach proposed by \WoCP seems promising, most of the contributors having mostly predictions corresponding to their actual type.
Only five human contributors have a higher number of \emph{bot} predictions than \emph{human} predictions. These contributors will be consistently but wrongly predicted as \emph{bot} by \WoCP.

To assess to which extent BoDeGHa can be improved by \WoCP, we evaluated both models on the 84 contributors. Table~\ref{table:misclass} reports on the resulting number of true positives (TP), true negatives (TN), false positives (FP), false negatives (FN) as well as on the accuracy (Acc), precision (Prec), recall and F1 scores of the two models.

\begin{table}[t]
	\centering
	\caption{Score comparison between BoDeGHa and \WoCP}
	\begin{tabular}{ r | r  r  r  r  r  r  r  r}
		 & TP &  TN & FP & FN & Acc & Prec & Recall & F1\\
		\hline
		BoDeGHa& 928 & 288 & 79 & 31 & 91.7 & 92.2 & 96.8 & 94.4\\
		\WoCP & 959 & 348 & 19 & 0 & 98.6 & 98.1 & 100.0 & 99.0 \\
	\end{tabular}
	\label{table:misclass}
\end{table}

We observe that \WoCP actually improves the predictions made by BoDeGHa.
\WoCP replaced a total of 101 predictions out of 1,326 (i.e., 7.6\%): 65 \emph{bot} predictions were correctly converted to \emph{human} predictions, while 36 \emph{human} predictions were converted to \emph{bot} predictions, among which 31 correspond to actual bots. This leads the number of false negatives to drop from 31 to 0, and the number of false positives to decrease from 79 to 19. These 19 incorrect predictions correspond to the five human contributors above the diagonal line in Figure~\ref{fig:bothum}.
As a consequence, \WoCP has higher accuracy, precision, recall and F1 scores compared to BoDeGHa.

\subsection*{RQ3: To which extent can we complete predictions?}

So far, we relied on the wisdom of the crowd principle, using the most frequent prediction to fix the less frequent predictions.
This question aims to determine whether a similar approach can be followed to fix \emph{unknown} predictions as well.

\begin{figure}[!h]
	\centering
	\includegraphics[width = \columnwidth]{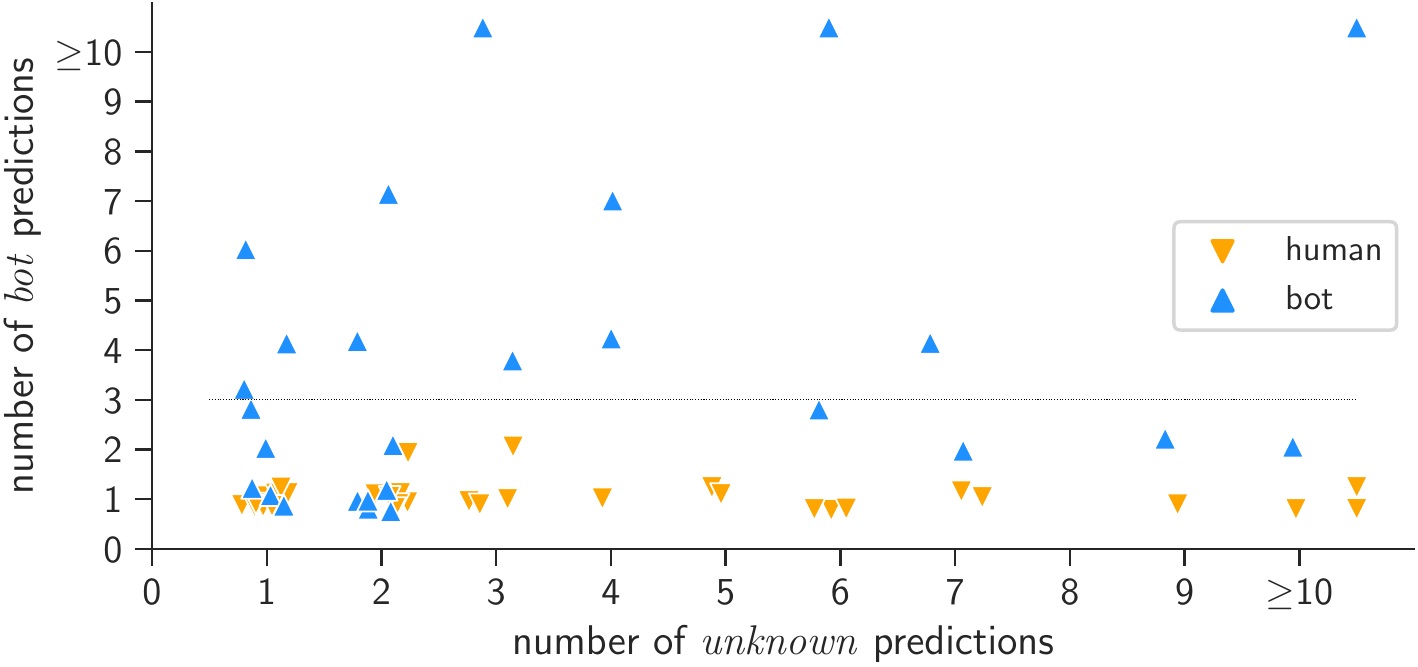}
	\caption{Accounts predicted as bot/unknown vs ground truth}
	\label{fig:botuk}
\end{figure}

Figure~\ref{fig:botuk} shows the number of \emph{unknown} and \emph{bot} predictions for the 63 contributors that were either predicted \emph{bot} or \emph{unknown} (i.e., that have no \emph{human} predictions).
We observe that the situation is more delicate than for $RQ2$.
Indeed, many actual human contributors are among the contributors having only \emph{bot} and \emph{unknown} predictions.
Converting the \emph{unknown} predictions to \emph{bot} predictions for these 33 human contributors would only increase the number of incorrect predictions for them.
For instance, while converting the 184 \emph{unknown} predictions of the 30 bots increases the number of correct predictions from 336 to 520, doing the same for the 158 \emph{unknown} predictions of the 33 human contributors increases the number of incorrect predictions from 35 to 193.

Nevertheless, we observe that most of these human contributors have a low number of \emph{bot} predictions compared to the actual bot contributors. For instance, there are 17 bots and no human having three or more \emph{bot} predictions. On the other hand, all human contributors and ``only'' 13 bots have one or two \emph{bot} predictions.
Converting only the \emph{unknown} predictions of contributors having three or more \emph{bot} predictions would increase the number of correct predictions from 318 to 460, without increasing the number of incorrect predictions.
However, since this threshold of ``3+ \emph{bot} predictions'' is obtained by observation, it cannot be integrated into the \WoCP model without prior validation on another dataset.\section{Conclusion}
\label{sec:conclusion}

Bots are being more and more used to automate some of the effort-intensive and repetitive activities that are part of the software development process in GitHub repositories.
Identifying bots is not only useful for researchers conducting socio-technical studies, but also for practitioners and funding organizations to identify contributors and to accredit them. BoDeGHa is one of the few tools that were proposed in the literature to detect bots in software repositories.
Although BoDeGHa works well at the repository level, it may still produce diverging or incomplete predictions for contributors being active in multiple repositories.
This paper gave preliminary insights on a novel approach based on the wisdom of the crowd to improve the performance of BoDeGHa.
Based on a dataset of around one thousand repositories, we investigated how frequently contributors are active in multiple repositories, how frequently BoDeGHa produces diverging or incomplete predictions, and to which extent these predictions can be fixed.
We found that around one third of contributors are active in the commenting activity of multiple repositories. We found that these contributors led BoDeGHa to produce a significant amount of diverging and incomplete predictions.
We proposed \WoCP, a new model relying on individual predictions made by BoDeGHa. Although BoDeGHa is already one of the best bot identification tools~\cite{botse_paper_2}, \WoCP improved its accuracy by 6.9\% for the considered cases. This suggests that relying on the wisdom of the crowd principle is a promising approach to improve bot detection models across repositories.
We also investigated to which extent such an approach can be used to fix incomplete predictions. The initial insights we obtained are encouraging, but they need to be validated on a different, larger dataset.

In a near future, we will consider extending BoDeGHa by integrating \WoCP to allow practitioners to benefit from this improved bot detection model.
We plan to evaluate and compare other approaches that exploit more data to detect bots, for example, by modifying the input and data processing part of BoDeGHa so that it can be applied on several repositories at once (i.e., aggregating comments from multiple repositories instead of aggregating predictions made from these repositories).
We also plan to evaluate to which extent other tools than BoDeGHa can benefit from the approach proposed in this paper.
\section*{Acknowledgement}
This work is supported by DigitalWallonia4.AI research project ARIAC (grant number 2010235), as well as by the Fonds de la Recherche Scientifique -- FNRS under grant numbers O.0157.18F-RG43 and T.0017.18.

\bibliographystyle{IEEEtran}
\bibliography{botsbiblio.bib}

\end{document}